\documentclass[a4paper,12pt]{article}
\pdfoutput=1 % if your are submitting a pdflatex (i.e. if you have
\usepackage{jcappub} % for details on the use of the package, please
\usepackage{fancyhdr}
\usepackage{fancybox}

\usepackage{color}
\usepackage{soul}
\usepackage{latexsym}
\usepackage{amsfonts}
\usepackage{feynmp}
\usepackage{graphicx,wrapfig,float,slashed,indentfirst}
\usepackage{amsmath,amssymb,epsfig,graphicx,xcolor}

\usepackage{tabu}
\usepackage{ragged2e}
\usepackage{lipsum}
\usepackage{siunitx}
\usepackage{ulem}
\allowdisplaybreaks
\usepackage{pifont}

\usepackage{tikz}
\usetikzlibrary{arrows,positioning,shapes}
\usetikzlibrary{decorations.pathmorphing}
\usetikzlibrary{decorations.markings}

\makeatletter
\def\@fpheader{\hfill\today}
\makeatother

\newcommand{\PreserveBackslash}[1]{\let\temp=\\#1\let\\=\temp}
\newcolumntype{C}[1]{>{\PreserveBackslash\centering}p{#1}}
\newcolumntype{R}[1]{>{\PreserveBackslash\raggedleft}p{#1}}
\newcolumntype{L}[1]{>{\PreserveBackslash\raggedright}p{#1}}

\def\tanb{\tan\beta}
\def\cotb{\cot\beta}
\def\sinb{\sin\beta}
\def\cosb{\cos\beta}
\def\sina{\sin\alpha}
\def\cosa{\cos\alpha}

\def\sbma{\sin(\beta-\alpha)}
\def\cbma{\cos(\beta-\alpha)}
\def\gam{\gamma}
\def\lam{\lambda}

\def\hl{h}
\def\hh{H}
\def\ha{A}

\def\hpm{H^\pm}

\def\mhl{m_{\hl}}
\def\mhh{m_{\hh}}
\def\mha{m_\ha}
\def\mhpm{m_{\hpm}}

\def\fnp{f_n \over f_p}

\def\gev{~{\rm GeV}}

\def\br{{\rm BR}}
\def\anti{\overline}

\def\ie{{\it i.e.}}

\def\omgs{\Omega_S h^2}
\def\hsm{{H_{\rm SM}}}

\def\Eq#1{Eq.~(\ref{#1})}

\def\beq{\begin{equation}}
\def\eeq{\end{equation}}
\def\bea{\begin{eqnarray}}
\def\eea{\end{eqnarray}}
\def\bit{\begin{itemize}}
\def\eit{\end{itemize}}
\def\ben{\begin{enumerate}}
\def\een{\end{enumerate}}

\def\zpr{\mathbb{Z'}_2}

\def\sigsi{\sigma^{\rm SI}}
\def\gam{\gamma}

\def\lam{\lambda}
\def\Lam{\Lambda}

\def\typeii{Type~II}

\def\mhpm{m_{\hpm}}

\def\lsim{\mathrel{\raise.3ex\hbox{$<$\kern-.75em\lower1ex\hbox{$\sim$}}}}
\def\gsim{\mathrel{\raise.3ex\hbox{$>$\kern-.75em\lower1ex\hbox{$\sim$}}}}

\def\ifmath#1{\relax\ifmmode #1\else $#1$\fi}
\def\ls#1{\ifmath{_{\lower1.5pt\hbox{$\scriptstyle #1$}}}}
\def\lss#1{\ifmath{^{\,\lower2.5pt\hbox{$\scriptstyle #1$}}}}

\def\tanb{\tan\beta}
\def\cotb{\cot\beta}
\def\sinb{\sin\beta}
\def\cosb{\cos\beta}
\def\sina{\sin\alpha}
\def\cosa{\cos\alpha}

\def\fnp{f_n/f_p}
\def\fn{f_n}
\def\fp{f_p}
\def\sigsip{\sigsi_{{\rm DM}-p}}

\title{Isospin-violating dark-matter-nucleon scattering via two-Higgs-doublet-model portals}
\author[a]{Aleksandra Drozd,}
\author[b]{Bohdan Grzadkowski,} 
\author[c]{John F. Gunion,}
\author[c,d,1]{Yun Jiang \note{Corresponding author.}}

\affiliation[a]{Theoretical Particle Physics and Cosmology Group, \\Physics Department, 
King's College London, London WC2R 2LS, UK}
\affiliation[b]{Faculty of Physics, University of Warsaw, Pasteura 5, 02-093 Warsaw, Poland}
\affiliation[c]{Department of Physics, University of California, Davis, CA 95616, U.S.A.}
\affiliation[d]{NBIA and Discovery Center, Niels Bohr Institute, University of Copenhagen, \\Blegdamsvej 17, DK-2100, Copenhagen, Denmark}

\emailAdd{Bohdan.Grzadkowski@fuw.edu.pl}
\emailAdd{gunion@physics.ucdavis.edu}
\emailAdd{yunjiang@nbi.ku.dk}

\abstract
{
We show that in a multi-Higgs model in which one Higgs fits the LHC $125\gev$ state, one or more of the other Higgs bosons can mediate DM-nucleon interactions with maximal DM isospin violation being possible for appropriate Higgs-quark couplings, independent of the nature of DM.  We then consider the explicit example 
of a \typeii\  two-Higgs-doublet model, identifying the $h$ or $H$ as the $125\gev$ state while the $H$ or $h$, respectively,  mediates DM-nucleon interactions. Finally, we show that if a stable scalar, $S$, is added then it can be a viable light DM candidate with correct relic density while obeying all direct and indirect detection limits.
}

\begin{document}
\maketitle

\flushbottom

\tikzstyle{every picture}+=[remember picture]
%%%%%%%%%%%%%%%%%%%%%%%%%%%%%%%%%%%%%%%%%%%%%%%%%%%%%%%%%%%%%%%%%%%%%%%%%%%%%%%%%%%%%%%%%%%%%%%%%%%%%%%%%%%%%%%%%%%%%%%%%%%%%%%%%%%%%%%%%%%%%%%
\pgfdeclaredecoration{complete sines}{initial}
{
    \state{initial}[
        width=+0pt,
        next state=sine,
        persistent precomputation={\pgfmathsetmacro\matchinglength{
            \pgfdecoratedinputsegmentlength / int(\pgfdecoratedinputsegmentlength/\pgfdecorationsegmentlength)}
            \setlength{\pgfdecorationsegmentlength}{\matchinglength pt}
        }] {}
    \state{sine}[width=\pgfdecorationsegmentlength]{
        \pgfpathsine{\pgfpoint{0.25\pgfdecorationsegmentlength}{0.5\pgfdecorationsegmentamplitude}}
        \pgfpathcosine{\pgfpoint{0.25\pgfdecorationsegmentlength}{-0.5\pgfdecorationsegmentamplitude}}
        \pgfpathsine{\pgfpoint{0.25\pgfdecorationsegmentlength}{-0.5\pgfdecorationsegmentamplitude}}
        \pgfpathcosine{\pgfpoint{0.25\pgfdecorationsegmentlength}{0.5\pgfdecorationsegmentamplitude}}
}
    \state{final}{}
}

\tikzset{
fermion/.style={thick,draw=black, line cap=round, postaction={decorate},
    decoration={markings,mark=at position .65 with {\arrow[black]{stealth}}}},
photon/.style={thick, line cap=round,decorate, draw=black,
    decoration={complete sines,amplitude=4pt, segment length=5pt}},
boson/.style={thick, line cap=round,decorate, draw=black,
    decoration={complete sines,amplitude=4pt,segment length=8pt}},
gluon/.style={thick,line cap=round, decorate, draw=black,
    decoration={coil,aspect=1,amplitude=3pt, segment length=8pt}},
scalar/.style={dashed, thick,line cap=round, decorate, draw=black},
ghost/.style={dotted, thick,line cap=round, decorate, draw=black},
->-/.style={decoration={
  markings,
  mark=at position 0.6 with {\arrow{>}}},postaction={decorate}}
 }

%%%%%%%%%%%%%%%%%%%%%%%%%%%%%%%%%%%%%%%%
\section{Introduction}
\vspace*{-.1in}

One of most outstanding failures of the Standard Model (SM) is the lack of a candidate for dark matter (DM), the latter constituting 27\% of the energy of the universe~\cite{Adam:2015rua}. Many models have been proposed for DM in a variety of beyond-the-SM theories. Higgs bosons could play an important role in two ways. First, one or more Higgs could mediate interactions between nucleons and DM.  Second, DM could itself be a  Higgs boson. In this letter, we consider a two-Higgs-doublet model (2HDM) within which there are two CP-even Higgs bosons, $h$ and $H$ ($\mhl<\mhh$), where one fits the SM-like state at $125\gev$.  We show that if the $h$ and $H$ mediate the interactions of DM with quarks we can arrange for the DM-nucleon interactions to be isospin-violating, thereby allowing light dark matter to be consistent with the LUX~(2016) limits~\cite{Akerib:2016vxi} at low DM mass, independent of the nature of the DM particle itself.  Next, we demonstrate that if the 2HDM is extended
to include a stable singlet scalar boson, $S$, whose interactions with quarks are mediated by the $h$ and $H$, we can choose parameters so that the $S$ can provide the observed relic density even for $m_S<60\gev$ without violating any theoretical or phenomenological constraints. 

The minimal SM extension (called xSM) for which DM might be a Higgs boson  is to add a scalar singlet field $S$ protected by a $\zpr$ symmetry under $S\to -S$, communicating with the SM via a $\lam S^2 H^\dagger H$ interaction~\cite{McDonald:1993ex,Burgess:2000yq}. However, to achieve correct relic DM abundance, $\omgs$, for $m_S\lsim 60\gev$ a rather large value of the portal coupling $\lambda$ is required. This leads to both too large $\br(\hsm\to SS)$ and a direct DM detection cross section exceeding the old LUX~(2013) upper limit~\cite{He:2011gc}.  

Both problems can be cured in the 2HDMS model~\cite{Drozd:2014yla} in which a real gauge-singlet scalar, $S$,  is added to the two doublet fields of the 2HDM. As above, if a $\zpr$ symmetry is imposed and we require that $S$ not have a vacuum expectation value (vev) then the $h$ and $H$ of the 2HDM will be mass eigenstates and the $S$ can be dark matter.
The main idea is that if the $h$ ($H$)  is identified as the $125\gev$ state (the $h125$ and $H125$ scenarios, respectively)
it can have a very small portal coupling to $S$ (and therefore small $SS$ branching ratio) while  correct relic abundance can be achieved via relatively strong interactions of the $H$ ($h$)  with the $S$. 
 
In addition to being able to achieve correct $\omgs$ for a light $S$  with small $SS$ branching ratio of the SM-like Higgs, in the 2HDMS model with \typeii\ Yukawa couplings one can avoid the LUX~(2016) exclusion bounds for low mass DM.  
The key point is that in \typeii\ models the couplings of the non-SM-like Higgs to up- and down-type quarks, and therefore to protons and neutrons are not the same, and, for appropriate parameter choices, can even have opposite sign leading to a very suppressed cross section for DM scattering off of a nucleus~\cite{He:2011gc,He:2013suk,Cai:2013zga,Wang:2014elb,Drozd:2014yla}.
 
The paper is organized as follows. In Sec.~II we briefly describe the current status of direct detection experiments and show how isospin-violating interactions of DM are possible in the \typeii\ 2HDM context, independently of whether or not dark matter is a Higgs boson. In Sec.~III we introduce the \typeii\ 2HDMS and find parameters for which the $S$ is a fully viable dark matter candidate. We end with a summary of our results. 

%%%%%%%%%%%%%%%%%%%%%%%%%%%%%%%%%%%%%%%%
\vspace*{-.1in}
\section{Direct detection of Dark Matter and isospin-violation \label{seci}}
\vspace*{-.1in}

DM is a compelling window to new physics and a primary means for its direct detection is via scattering off nucleons. 
Experimental results are  typically translated into the event rate (or limit)  for the spin-independent cross section for DM scattering off a nucleon $\sigsi_{{\rm DM}-N}$ as a function of DM mass. The strongest  exclusion limits are currently those from  LUX~\cite{Akerib:2013tjd} and, in the very-low mass regime (\ie\ DM mass below $15\gev$), SuperCDMS~\cite{Agnese:2014aze}.

Translating from experimental data to $\sigsi_{{\rm DM}-N}$ involves many assumptions, including use of the Standard Halo Model (as in~\cite{Akerib:2013tjd}) and elastic scattering at zero-momentum transfer with a short range contact interaction.  In particular, limits on $\sigsi_{{\rm DM}-N}$ are typically given assuming that DM couples equally to the neutron and proton, the strengths of these couplings being denoted by $\fn$ and $\fp$ --- see \cite{Drozd:2014yla}  for details using our conventions.  
If $\fnp \neq 1$, one must apply a rescaling factor $\Theta_X$ to convert the predicted DM-proton cross-section $\sigsip$ to the DM-nucleon cross section $\sigsi_{{\rm DM}-N}$ obtained assuming $\fnp=1$:
\beq
\sigsi_{{\rm DM}-N} = \sigsip \, \Theta_X(f_n,f_p),
\eeq
where the rescaling factor $\Theta_X$ for a multiple isotope detector is defined in~\cite{Drozd:2014yla}.
When $f_n/f_p \neq 1$, $\Theta_X(f_n,f_p)$ will depend upon the isotope abundances (which are detector-dependent) and can be as small as $\sim 10^{-4}$ when $\fnp$ is close to $-1$, $-0.8$, $-0.7$ for target nucleons Si, Ge, and Xe, respectively, (with weak dependence on $m_S$)~\cite{Feng:2013fyw}.

\begin{figure}[t]
\begin{center}
\includegraphics[width=0.55\textwidth]{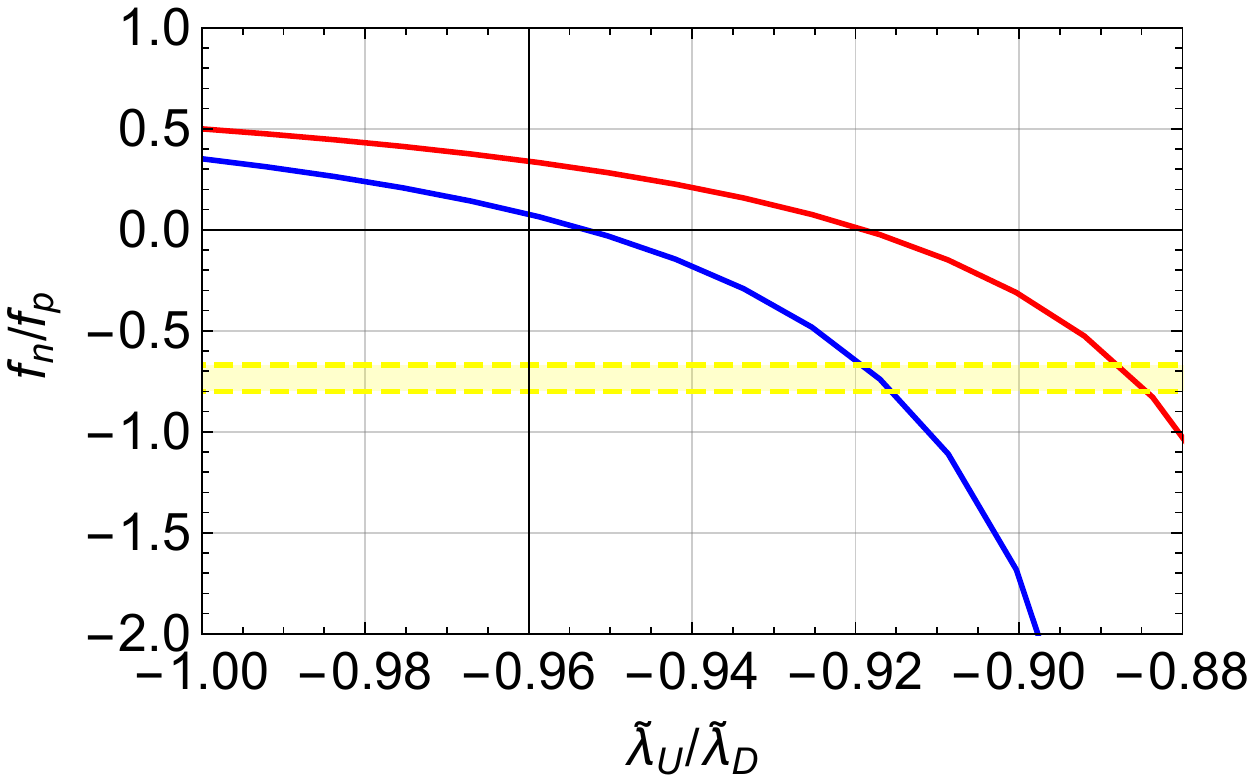}
\end{center}\vspace*{-5mm}
\caption{
The correlation between $\fnp$ and $\tilde\lam_{U}/\tilde\lam_{D}$ with (red) and without (blue) QCD NLO corrections and using $f^n_{Tu}=0.011, f^n_{Td}=0.0273, f^n_{Ts}=0.0447$  and $f^p_{Tu}=0.0153, f^p_{Td}=0.0191, f^p_{Ts}=0.0447$, all as employed in micrOMEGAs \cite{Belanger:2013oya}.  
The yellow band  corresponds to $\fnp$ in the range -0.67 to -0.8.} 
\label{fnp-rCUD}
\vspace{-.1in}
\end{figure}

As we now describe,  such $\fnp$ values can be achieved in multi-Higgs models, independently of the nature of DM. One Higgs must be identified with the SM-like state at $125\gev$ and have very weak coupling to DM, while one or more of the other Higgs bosons should be primarily responsible for mediating DM-quark interactions.
As derived in~\cite{Drozd:2014yla}, the general expression for $\fnp$  is
\beq
\frac{f_n}{f_p} ={m_n \over m_p} {F^n_u \tilde\lam_{U} + F^n_d \tilde\lam_{D} \over  F^p_u \tilde\lam_{U} + F^p_d \tilde\lam_{D} }
\label{fnp}
\eeq
where 
\bea
F^N_u &=& f^N_{Tu} + \sum_{q=c,t}{2\over 27} f^N_{TG} \left(1+{35 \over 36 \pi} \alpha_S (m_q)\right) \label{FNu} \\
F^N_d &=& f^N_{Td} +  f^N_{Ts} + {2\over 27} f^N_{TG} \left(1+{35 \over 36 \pi} \alpha_S (m_b) \right) \label{FNd}
\eea
($N=p,n$) and the scale-dependent $\alpha_S$ terms account for the QCD NLO corrections (not included in~\cite{Drozd:2014yla})
while $f^N_{TG} = 1-\sum_{q=u,d,s}f^N_{Tq}$.
$\tilde\lam_{U}$ and $\tilde\lam_{D}$ are defined as follows
\beq
 \tilde\lam_{U} =  \sum_{\mathcal{H}} {\Lam_{\mathcal{H}} \over m_{\mathcal{H}}^2} C^{\mathcal{H}}_U, \quad\quad  \tilde\lam_{D} = \sum_{\mathcal{H}} {\Lam_{\mathcal{H}} \over m_{\mathcal{H}}^2} C^{\mathcal{H}}_D \,,
\label{lamU}
\eeq
where  $\sum_{\mathcal{H}}$ sums over the Higgs mediators contributing to the $t$-channel diagrams,  $C^{\mathcal H}_{U,D}$ denote the $\mathcal{H}$ couplings to up-,  down-type quarks, respectively, normalized to their SM values, while the $\Lambda_{\mathcal H}$ are dimensionless parameters specifying the strengths of the ${\mathcal H}$ couplings to a pair of DM particles. Fig.~\ref{fnp-rCUD} shows the ratio $\fnp$ as a function of $\tilde\lam_{U} /\tilde\lam_{D} $.
A negative value of $\fnp$ is obtained in a narrow range of $\tilde\lam_{U}/\tilde\lam_{D}$ around $-0.9$. The exact $\fnp$ value is very sensitive to  the QCD corrections. The choice which gives  maximal suppression for Xe as well as maximal relative scaling between Xe and Si is  $\fnp \simeq -0.7$, which occurs at $\tilde\lam_{U}/\tilde\lam_{D} \simeq -0.89$ and $-0.92$ when the QCD NLO correction is included or not, respectively \footnote{The possible role of NLO/multi-particle interactions in determining the precise $\fnp$ value needed to minimize Xenon dark-matter scattering rate is discussed in \cite{Cirigliano:2012pq}.}.

%%%%

\begin{table}[h]
\vspace{-.2in}
\caption{Tree-level vector boson couplings $C_V$ ($V=W,Z$) and fermionic couplings $C_{F}$ ($F=U,D$)
normalized to their SM values for the Type~II 2HDMs. }
\label{tab:couplings}
\begin{center}
\begin{tabular}{|c|c|c|c|}
\hline
Higgs & $C_V$ & $C_U$ & $C_D$  \cr
\hline
 $h$ & $\sin(\beta-\alpha)$   &  $\cosa/\sinb$ & $-{\sina/\cosb}$   \cr
\hline
 $H$ & $\cos(\beta-\alpha)$  &  $\sina/ \sinb$ & $\cosa/\cosb$ \cr
\hline
 $A$ & 0  & $\cotb$  & $\tanb$ \cr
\hline 
\end{tabular}
\end{center}
\vspace{-.15in}
\end{table}

\begin{figure}[t]
\begin{center}
\includegraphics[width=0.5\textwidth]{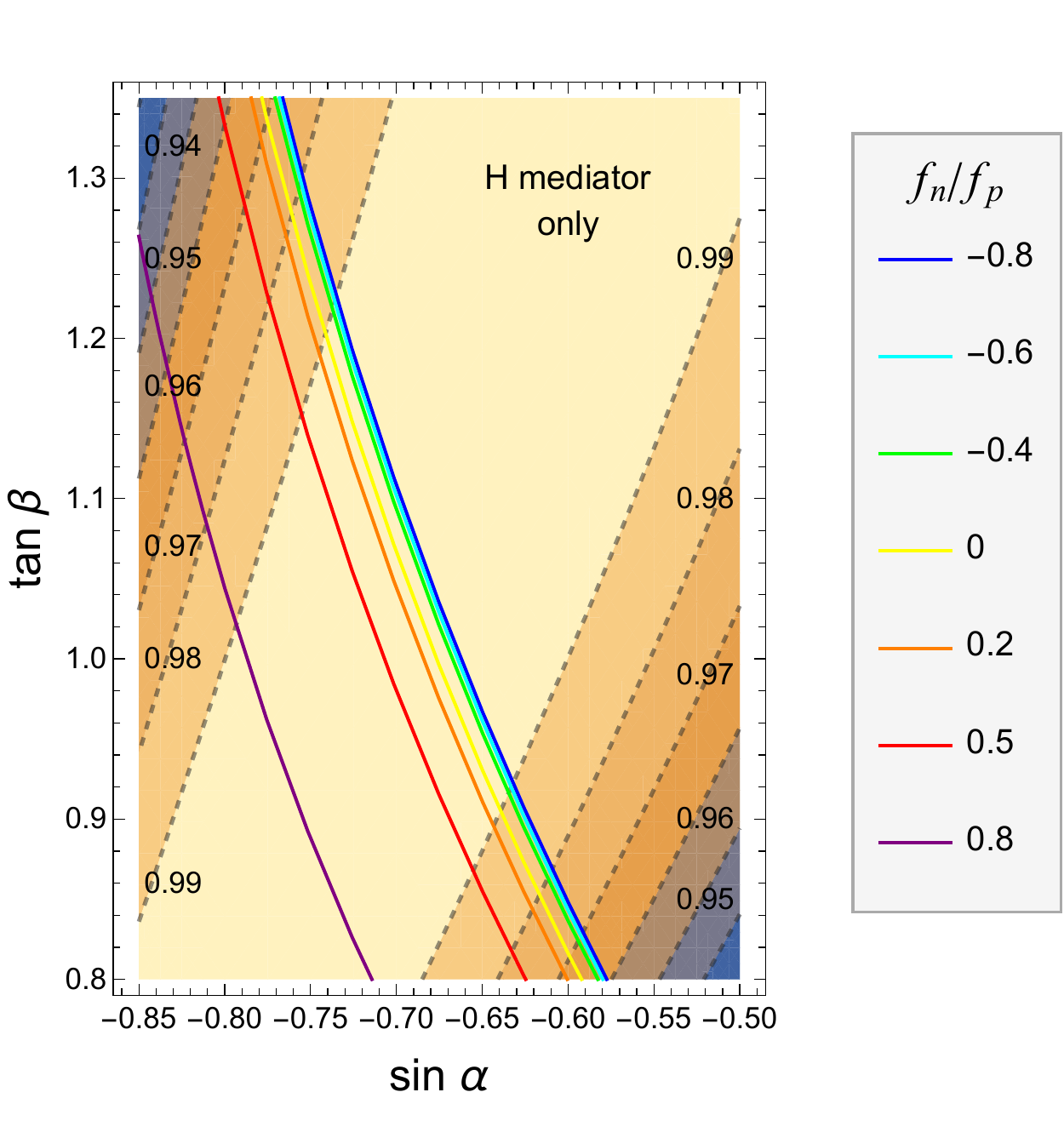}
\hspace*{-18mm}
\includegraphics[width=0.5\textwidth]{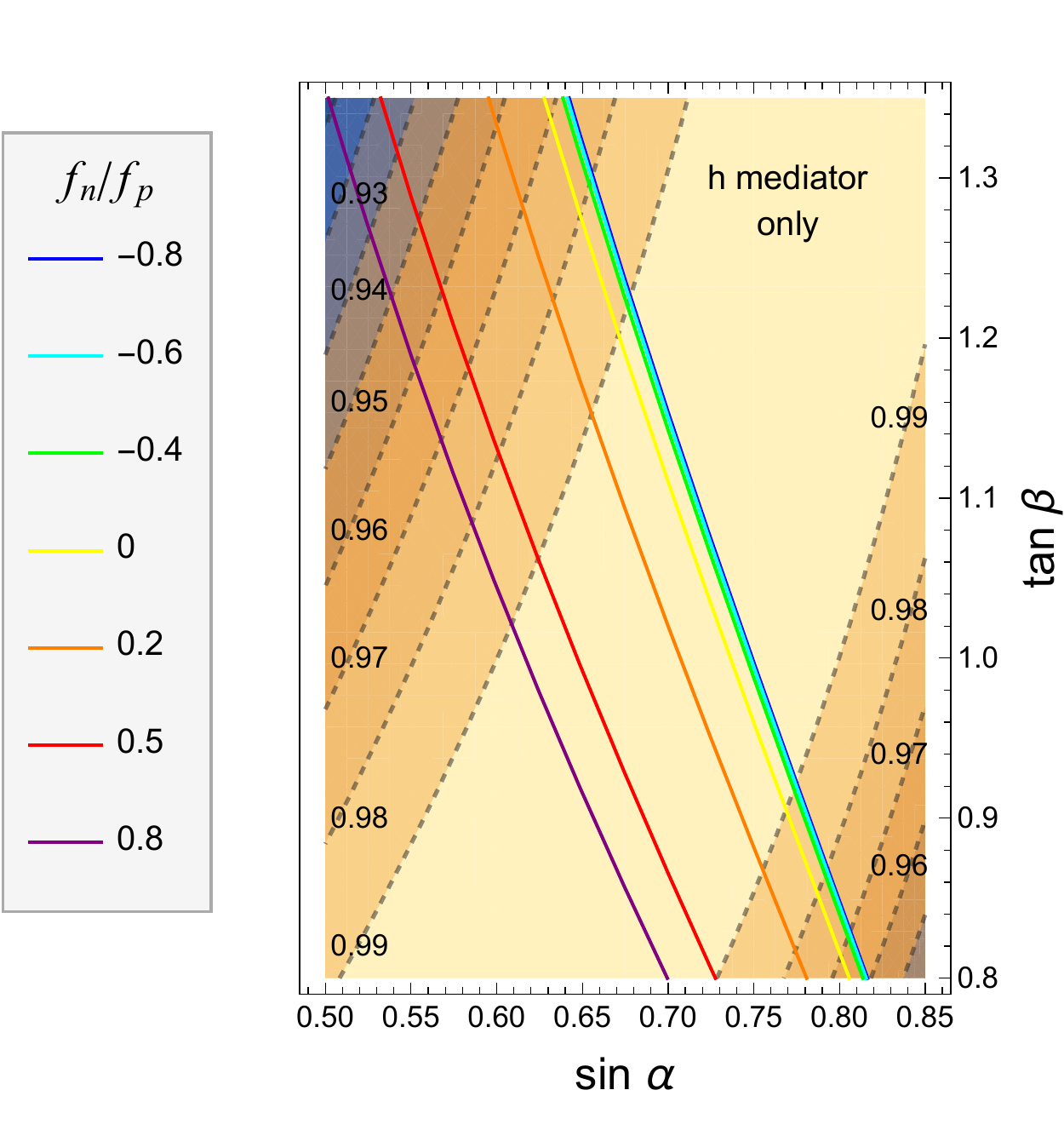}
\end{center}\vspace*{-7mm}
\caption{The left and right panels show contour plots (solid lines) of constant $\fnp$ in the ($\tanb,\sina$) space for the case
$m_{h}\sim 125\gev$ ($H$ is the mediator) and $m_{H}\sim 125\gev$ ($h$ is the mediator), respectively.
NLO QCD corrections are taken into account. The dashed lines are contours of constant $\sbma$ and $\cbma$ in left and right panels,
respectively.}
\label{tb-sa-fnp}
\vspace{-.1in}
\end{figure}

The key ingredient in achieving $\tilde\lam_U/\tilde \lam_D\sim -0.9$ is that the Higgs mediators have appropriately different couplings to up and down quarks.  A 2HDM of \typeii\ is such a model. Using \Eq{fnp} and the Higgs-quark couplings $C_U,C_D$ of Table~\ref{tab:couplings}, a given value of  $\fnp$ requires:
\beq
{\tan\beta = -{(\fnp) F^p_u - (m_n/ m_p) F^n_u \over (\fnp) F^p_d - (m_n / m_p) F^n_d }  {w+\tan\alpha \over 1-w \tan\alpha}}
\eeq
where $w={\Lam_h \over \Lam_H} {m_H^2 \over m_h^2}$. 
Requiring that the SM-like Higgs has zero coupling to a pair of DM particles so as to avoid its having invisible decays, implies $w \to 0$ ($w \to \infty$) for the $h125$ ($H125$) scenario.
In Fig.~\ref{tb-sa-fnp}, we plot $\tanb$ versus $\sina$ in these two cases for various values of $\fnp$. The value of $\fnp \sim -0.7$ needed to suppress Xe limits corresponds to the very narrow band between the solid blue and cyan lines.
In the figures, we also show (dashed) lines of constant $C_V^h=\sbma$ ($C_V^H=\cbma$) in the left (right) panels. Requiring $C_V^h\sim 1$  ($C_V^H\sim 1$) for the $h125$ ($H125$) to be very SM-like implies that $\tanb$ and $\sina$ must lie within the broad central yellow band. Combining this with the $\fnp\sim -0.7$ requirement leaves only a small region in each of the $(\tanb,\sina)$ parameter spaces, 
located near $\tanb\sim 1$ and 
$\sin\alpha \sim -0.7~(+0.7)$, implying $C_D^H\sim -C_U^H\sim1$ ($C_U^h\sim -C_D^h\sim1$) for the $h125$ ($H125$) scenario. 

%%%%%%%%%%%%%%%%%%%%%%%%%%%%%%%%%%%%%%%%%%%%%%%%%%%%%%%%%%%%%%%%

%%%%%%%%%%%%%%%%%%%%%%%%%%%%%%%%%%%%%%%%%%%%%%%%%%%%%%%%%%%%%%%%%

\vspace*{-.1in}
\section{The 2HDMS Dark-Matter Model}
\vspace*{-.1in}

Let us now consider the 2HDMS model in which a singlet scalar Higgs, $S$, is added to the 2HDM. The $\zpr$ symmetric and gauge-invariant 2HDMS scalar potential was given in  \cite{Grzadkowski:2009iz} and \cite{Drozd:2014yla}.
In the end, the terms associated with the $S$ in the potential of importance to this study are:
{\small
\bea
V_S &\!\!=\!\!& {m_S^2 \over 2} S^2 +v \left( \lam_{h} h + \lam_{H}  H \right) S^2 +   \lam_{H^+H^-}H^+H^- S^2
+ {\lam_S \over 4!}S^4 \cr
&&  + \left( \lam_{hh}hh + \lam_{hH}hH +\lam_{HH}HH  +   \lam_{AA}AA  \right) S^2 \, .
\label{V_S}
\eea
}
(The previously employed generic portal couplings appearing in \Eq{lamU} are given by $\Lambda_{h,H}=-2\lam_{h,H}$.)

Because its interactions are invariant under  $S\to -S$, the $S$ can be DM provided it does not acquire a vev.  Further, the $S$ does not affect the fits of \cite{Dumont:2014wha,Bernon:2014nxa} to the LHC Higgs data so long as the 2HDM state of mass $125\gev$ has small branching ratio to $SS$ pairs.  To avoid such decays we require $\lam_h=0$ or $\lam_H=0$ in the $h125$ or $H125$ scenarios, respectively.
For our numerical work, we employ the $\mhl=125\gev$ or $\mhh=125\gev$ parameter points of~\cite{Dumont:2014wha,Bernon:2014nxa} that described the LHC Higgs data at the (rather stringent) 68\% CL, supplemented by the latest $b\to s\gam$ constraint of $\mhpm\gsim 480\gev$ for the \typeii\ model~\cite{Misiak:2015xwa}. For each such point,  we scan over the independent singlet-sector parameters ($m_S$ and $\lam_H$ or $\lam_h$, respectively, fixing $\lam_S=2\pi$) and accept only points that satisfy perturbativity, tree level vacuum stability, tree level unitarity and  for which a proper electroweak vacuum is achieved. We also require that the precision electroweak $S$ and $T$ parameters fall within $\pm 3\sigma$ of their observed values.

Dark matter relic abundance, $\omgs$, is determined by the total DM annihilation rate. The relevant processes depend upon whether we consider the $h125$ or $H125$ scenario. 
For the $h125$ scenario, the amplitude diagrams for light dark matter ($m_S \leq 50 \gev$) are $SS\to H\to f\anti f$, $SS\to H\to \gam\gam$, and (relevant for $\mha\lsim m_S$) $SS\to H\to AA$ and $SS\to AA$ via contact interaction. 
In the $H125$ scenario the $SS$ annihilation tree-level diagrams are $SS\to h \to f\anti f$, $SS\to h\to \gam\gam$, $SS\to h \to hh$, $SS\to hh$ via $t,u$-channel $S$ exchange and via contact interaction. ($SS\to AA$ annihilation does not occur since $m_A>420\gev$ and  the $hh$ final states do not contribute unless $m_S\geq\mhl$.)  Also note that the parameter constraints needed to avoid large $\br(h\to AA)$ ($\br(H\to hh)$) when $m_A<\mhl/2~(\mhl<\mhh/2)$ in the $h125$ ($H125$) scenarios were studied in  \cite{Bernon:2014nxa} and are incorporated in our 2HDM fits --- they cause some variations of the phenomenology with $m_A$ ($m_h$). For example, in the $h125$ case  if $m_A<\mhl/2$ then correct $\omgs$ cannot be obtained if $m_A\geq m_S$, whereas if $\mha>\mhl/2$ then $\mha>m_S$ for the range of $m_S$ we consider and correct $\omgs$ is easily obtained.
Finally, we note that in the $H125$ case if $\mhl \sim 2m_S$   then s-channel $\hl$ exchange processes are strongly enhanced due to a resonance effect, whereas in the $h125$ case $\mhh\sim 2m_S$  is not possible.  

Thus, the main free parameter that determines $\omgs$ in the $h125$ ($H125$) scenarios is $\lam_H$ ($\lam_h$).
As studied in \cite{Drozd:2014yla}, for any  2HDM parameter point accepted by the analysis of \cite{Dumont:2014wha,Bernon:2014nxa}  it is straightforward to find singlet-sector parameter choices for which the observed relic density lies within the $\pm3\sigma$ window, $\omgs=  0.1187 \pm 0.0017$,  after satisfying all the theoretical and experimental constraints related to the Higgs sector.  (This is in sharp contrast to the xSM model mentioned in the Introduction.)  In the figures to follow, only points that have $\omgs$ in the above band (``correct" $\omgs$) are shown.

%%%%%%%%%%%%%%%%%%%%%%%%%%%%%%%%% 
\subsection {Collider bounds from direct searches for Higgs bosons}
%%%%%%%%%%%%%%%%%%%%%%%%%%%%%%%%%
Having taken into account the theoretical constraints on the 2HDMS and found parameter space points such that the $S$ state in this model constitutes dark matter producing the entire thermal relic abundance, we now turn to bounds from searching for non-SM Higgs at the LHC. First of all, the available bounds for Higgs masses below $62.5\gev$ were fully implemented in~\cite{Bernon:2014nxa}. Since we employ the points from that paper in the present work, these bounds are automatically taken into account. In this section we shall discuss additional collider bounds that must be imposed coming from searches for light Higgs bosons in the mass range of $62.5-125\gev$ at the LHC. 

Possibly relevant direct Higgs production searches are:
i) CMS~\cite{Aad:2014vgg} and ATLAS~\cite{Khachatryan:2014wca} limits on a light Higgs with mass $\gtrsim 90\gev$ decaying to $\tau\tau$ produced via gluon fusion or via $b\bar b$ associated production; 
and ii) CMS~\cite{Khachatryan:2015baw} limits on a light pseudoscalar Higgs boson of mass $25-80$ GeV produced in association with $b\bar{b}$ and decaying to a pair of $\tau$
leptons.~\footnote{Direct computation reveals that the cross section for a light $A$ and for a light $h$ are very similar in magnitude in $b\bar b$ associated production (and gluon fusion). Thus, the limits of~\cite{Khachatryan:2015baw} are, in principle, relevant for H125 scenario in which there is a light $h$ present.}
We find that these two constraints do not eliminate any of our points due to the fact that the predicted cross sections are about 1-3 orders in magnitude below the experimental limits.

Next, we examine the consistency of our model points with the recent CMS result~\cite{Khachatryan:2016are} on the search for a new heavy resonance decaying to a $Z$ boson and a light resonance ($h$ in our case), followed by $ Z \to \ell \ell$ and the light resonance decaying to $b\bar b$ or $\tau\tau$. In our model, these limits apply to the process $A\to Zh \to \ell\ell h $ with $h\to b\bar b $ or $\tau\tau$. 
Unfortunately, since the experimental limits are given using overlapping color coding, it is not possible to use the experimental plots to get precise limits as a function of $m_h$ and $m_A$. However, it is possible to extract the weakest and the strongest upper bounds at any given $(m_h,m_A)$. For $m_h<62.5\gev$, we find that the strongest (weakest) bound in the $b\bar b$ final state is 30~fb (100~fb) with corresponding bounds of 10~fb (100~fb) in the $\tau\tau$ final state. For $m_h>62.5\gev$, the strongest (weakest) bound in the $b\bar b$ final state is 10~fb (30~fb) with corresponding bounds of 3~fb (10~fb) in the $\tau\tau$ final state. It is convenient to compare these bounds to our model predictions by dividing the bound by the cross section for $gg\to A$ predicted in our model (taking $\tan\beta=1$, as appropriate in our model).  This gives us the ``weak" and ``strong" bounds on $\br(A\to Zh \to \ell\ell  b\bar b /\tau\tau)$.  

\begin{figure}[t]
\centering
\includegraphics[width=0.5\textwidth]{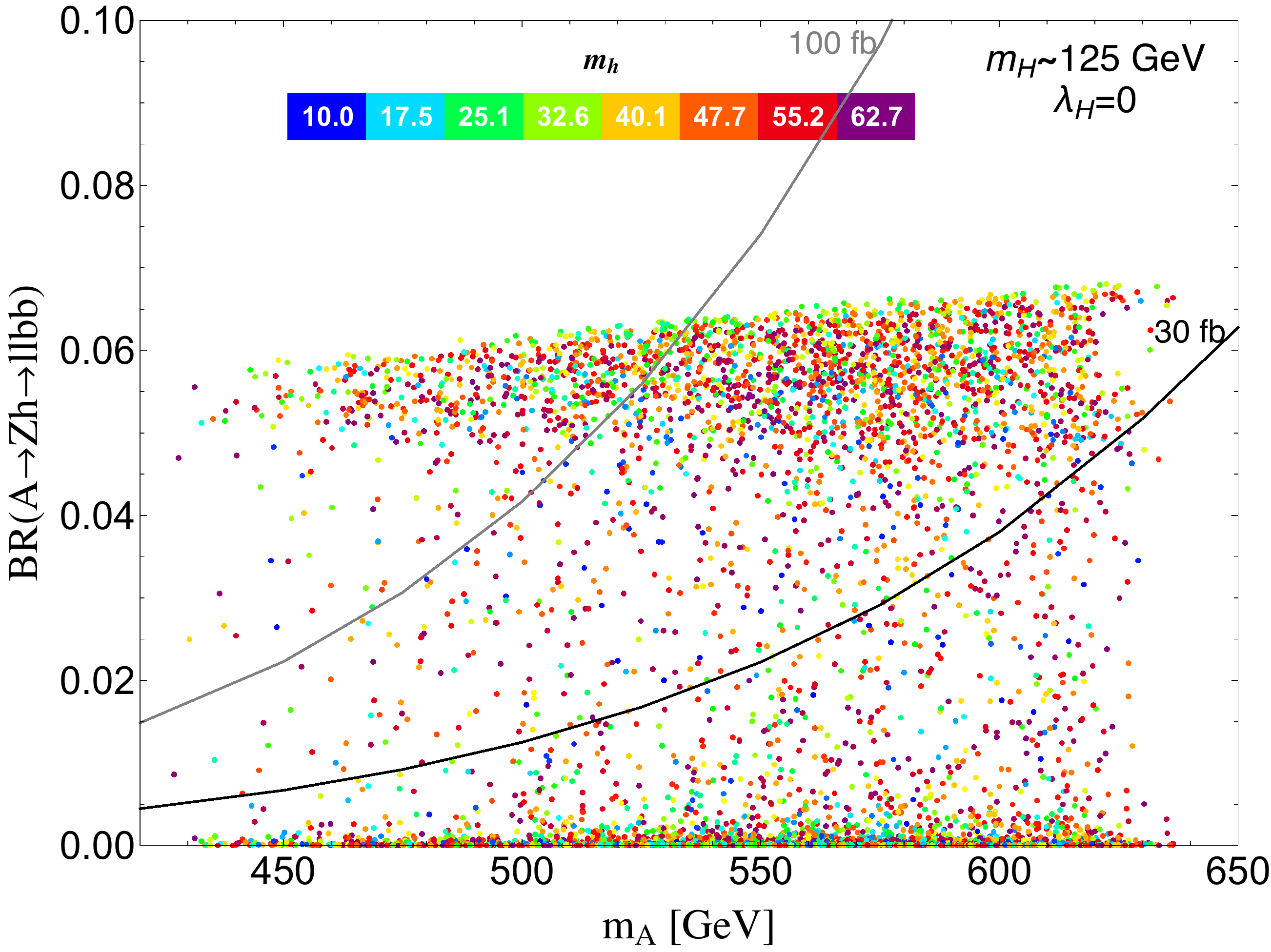}
\hspace{-3mm}
\includegraphics[width=0.5\textwidth]{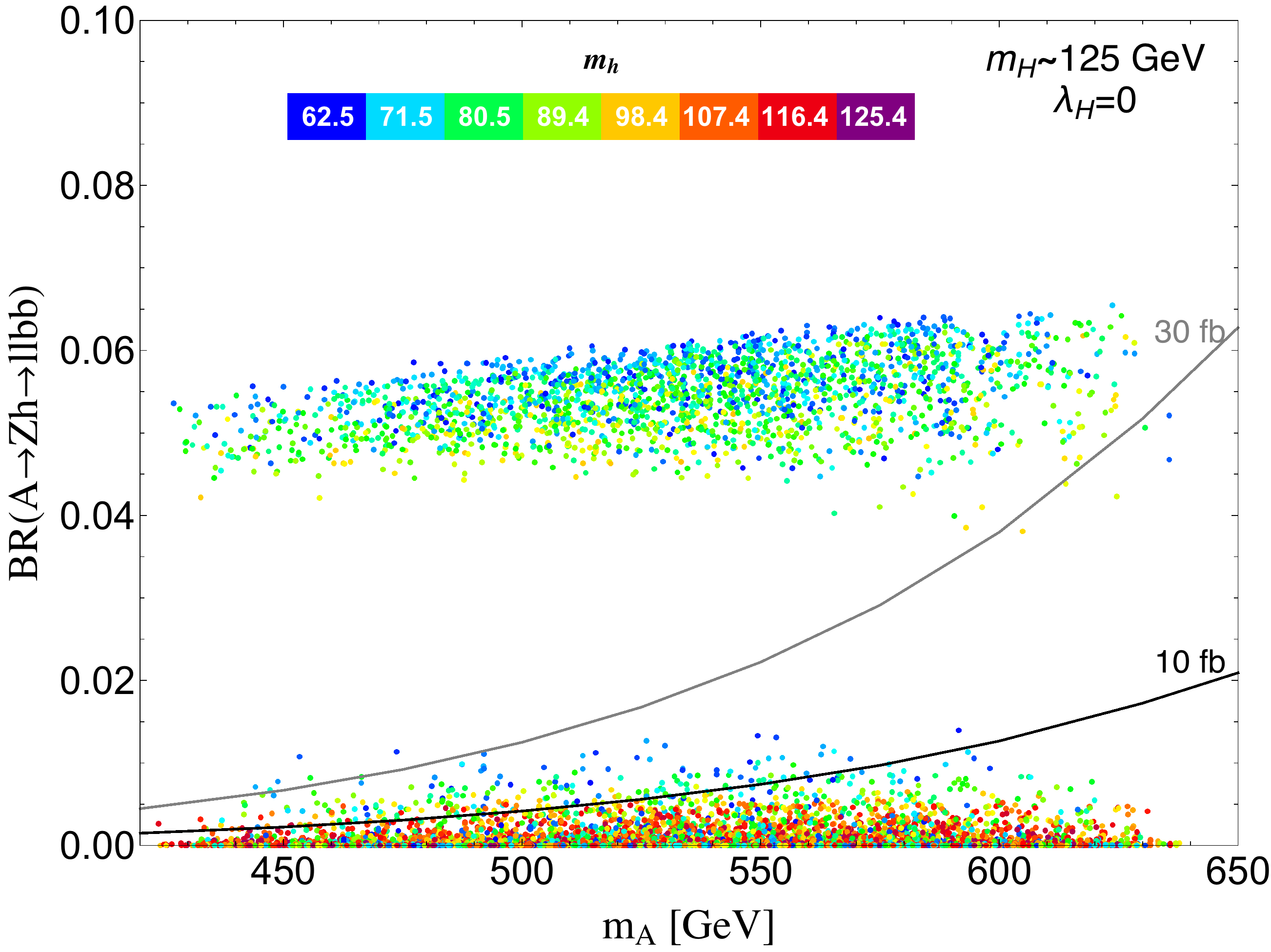}
\includegraphics[width=0.5\textwidth]{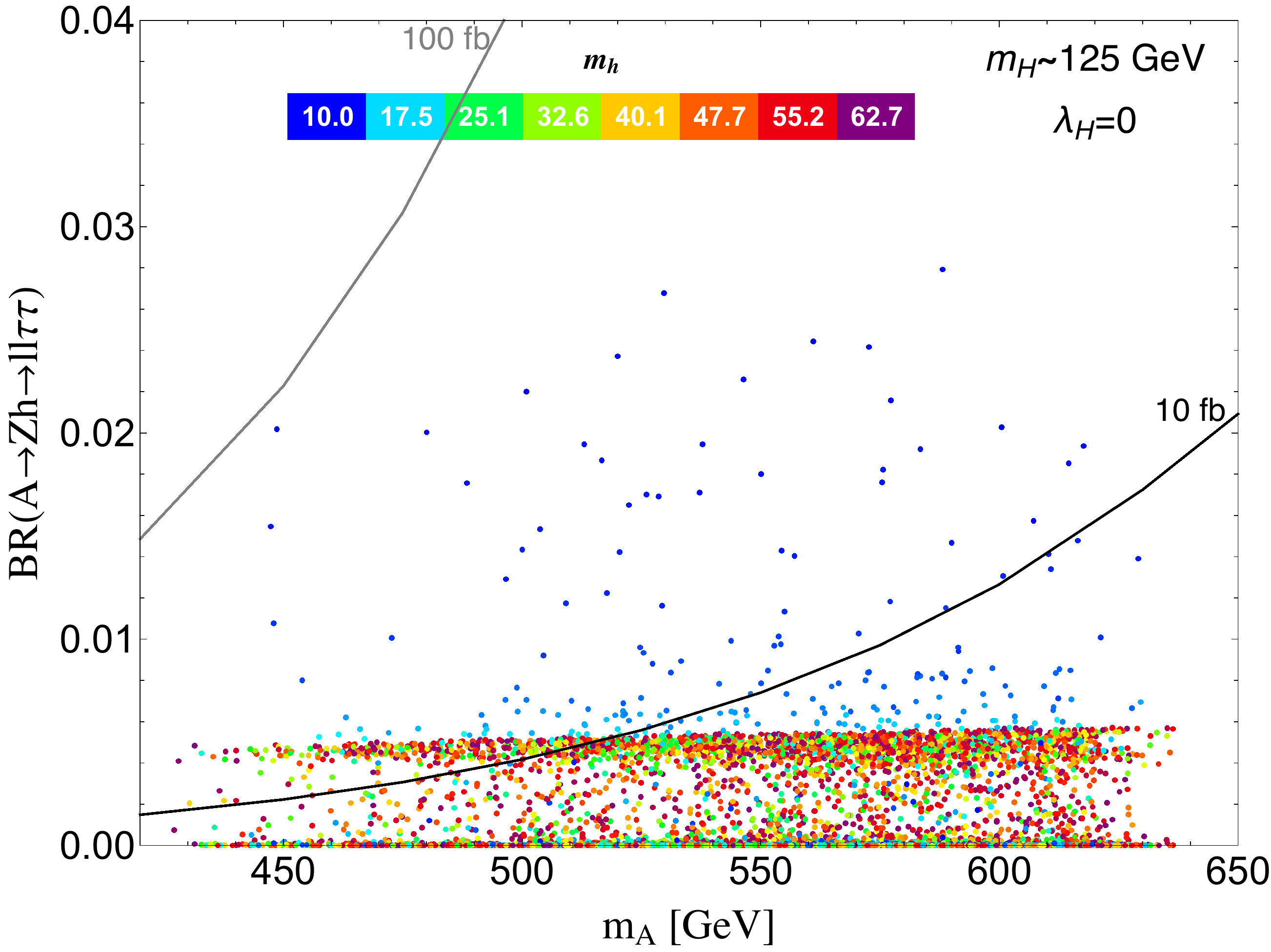}
\hspace{-3mm}
\includegraphics[width=0.5\textwidth]{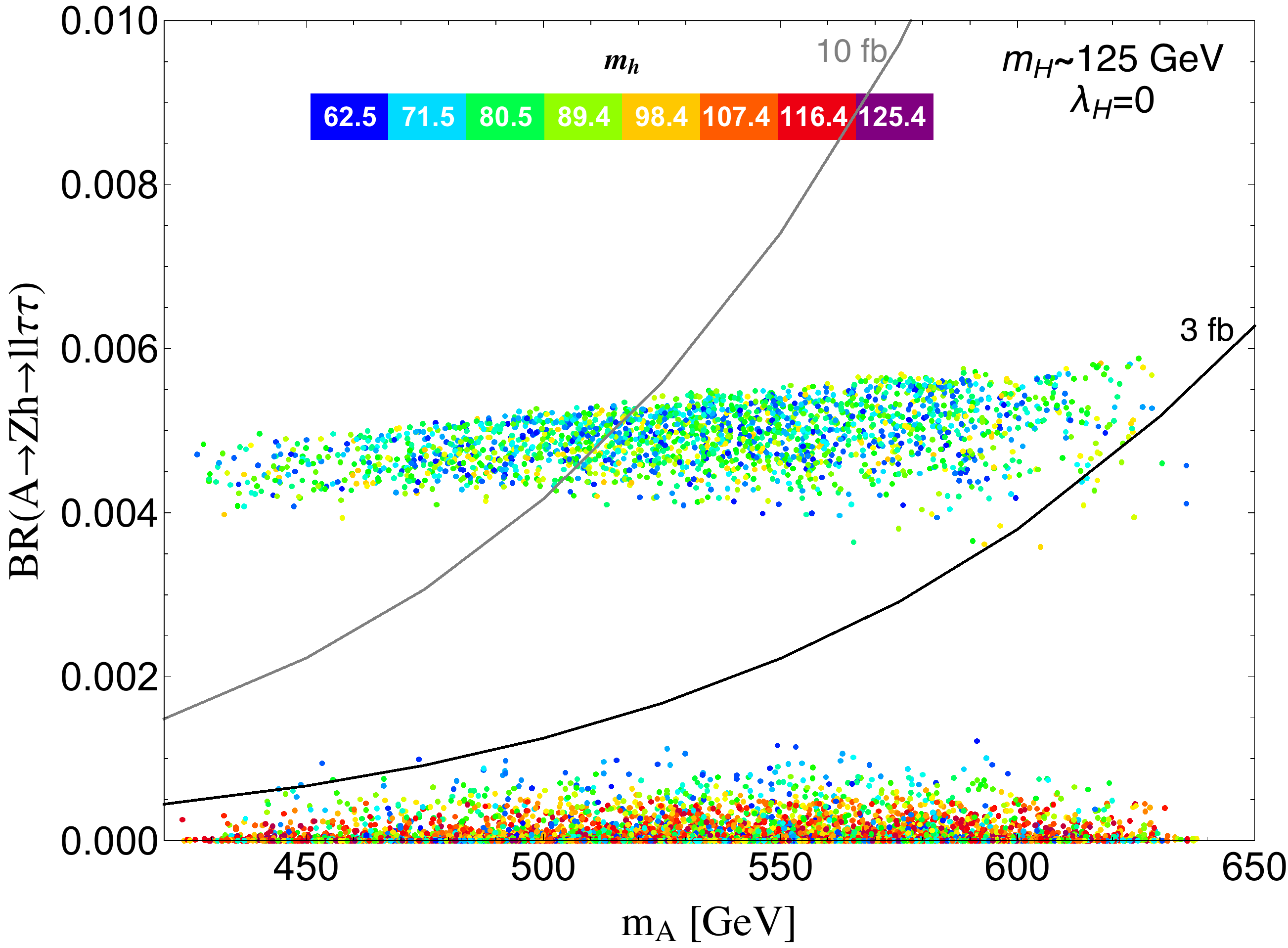}
\vspace{-3mm}
\caption{$\br(A\to Zh\to llbb)$ (upper) and $\br(A\to Zh\to ll\tau\tau)$ (lower) as a function of $m_A$. The point color indicates the value of $m_h$, which is below 62.5 GeV (left) and in the range of 62.5--125 GeV (right). We display the (strong and weak --- see text) experimental upper limits on the these branching ratios assuming $A$ production via gluon-fusion, where for each point the branching ratio limit is computed by dividing the cross section limit by $\sigma(gg\to A)$ as computed taking $\tan\beta=1$. In this scenario the heavy Higgs $H$ is identified as the SM-like state at 125 GeV. All points shown can produce correct relic abundance and are not excluded by the LUX~(2016) limit. Points below the black (gray) curve satisfy the strong (weak) constraints from non-SM-like Higgs searches at the LHC.}
\label{fig:bbtautaubound}
\end{figure}

In Fig.~\ref{fig:bbtautaubound}, the points show our predicted $\br(A\to Zh \to ll b\bar b /\tau\tau)$ as a function of pseudoscalar mass $m_A$, with coloring according to the $m_h$ value. The results are displayed separately for two different $m_h$ ranges relevant for  the subsequent discussion. 
In order to visualize the impact of the experimental data on our model, both the strong (minimal) and weak (maximal) upper limits on the cross section are shown by the black and gray curves in each plot, respectively.    The points above the black (gray) curves would be excluded by the strong (weak) limits.  
We see that the strong upper limit in the $b \bar b$ final state removes many points with $m_h > 30\gev$ and thus significantly constrains the light Higgs $h$ in the $H125$ scenario.  In particular, for the case $m_h>62.5\gev$, this limit entirely eliminates the points in the bulk with $\br(h\to b \bar b)\geq 70\%$,  pushing this branching ratio down to the 20\% level.
The $\tau\tau$ limits also have an impact for the $m_h \leq 30\gev$ points as shown in the lower left plot, while their importance becomes marginal below $15\gev$. For the purpose of showing the consistency of our model, we choose to adopt the ultra-conservatiove approach of imposing the strong upper limits --- that is in subsequent plots and discussions we retain only the points below the black solid curves. 

%%%%%%%%%%%%%%%%%%%%%%%%%%%%%%%%%
\subsection {Collider bounds from jet plus missing energy final states}
%%%%%%%%%%%%%%%%%%%%%%%%%%%%%%%%%
In this subsection, we consider the bounds from mono-jet+$\slashed {E}_T$ searches for dark matter.  Such searches have been performed by the LHC experimental groups~\cite{Khachatryan:2014rra,Aad:2015zva}, although, to date, results from Run~2 are only available from ATLAS~\cite{Aaboud:2016tnv}. Unfortunately, they present their results under assumptions that do not apply to our model.  Most critically, the effective operator approach is adopted to present the results. However, since the $\slashed {E}_T$ cuts employed are of order $100$'s of GeV,  the energy transfer in the collision exceeds the mediator ($h$) masses of interest to us and the effective operator approach is very inaccurate. In addition, our values of $m_h$ are such that the $h$ is mainly produced on-shell.  Thus, we do not think that the bounds presented in the experimental papers can be applied to our analysis.

Generally speaking, constraints from mono-jet+$\slashed {E}_T$ searches will be applicable to our scenarios when properly analyzed. In our model,   the $h$ mediator is mainly produced on-shell and will yield $\slashed {E}_T$ if the dark matter mass $m_S$ is below $m_h/2$ and if $h\to SS$ decays are dominant, as is the case  for many of our scan points (but not all). 
When the narrow width approximation is applicable, the jet+$\slashed {E}_T$ cross section  is the same as for jet+$h$ times $\br(h\to SS)$. For the bulk of our points, $\br(h\to SS) \simeq 100\%$ implying little dependence of the cross section on the nature of DM. Further, if the $\slashed {E}_T$ cuts are large, the implied jet energy will also be much larger than $m_h$ and the jet+$h$ cross section will then depend weakly on $m_h$, rising only slowly as $m_h$ decreases. For these reasons, the scalar-mediator results from the phenomenological analysis of~\cite{Harris:2014hga} are applicable to our model despite the fact that they assume fermion dark matter and only consider mediator masses above $125\gev$. Their Figure 5 displays, as a function of $m_h$, the limit on the ratio, defined as $\mu$,  of the cross section for jet+$\slashed {E}_T$ relative to that predicted if the mediator couplings to SM fermions have SM values. In the narrow width approximation, this ratio is equal to the ratio of the mediator-coupling-squared to SM particles relative to SM strength.  In their figure, only values of mediator mass, $m_h$, above $125\gev$ are plotted, for which the limit is of order $2-4$, falling slowly as $m_h$ decreases.  Extrapolation to lower $m_h$ values  suggests that it would only fall below 1 for $m_h$ values below $20-30\gev$ and maybe not even then.  Since our $h$ couplings are such that $\mu=1$ is predicted,  we conclude that the experimental limits from the 8~TeV, Run~1 data (and the projected limits from the 14~TeV, Run~2 data) do not exclude our preferred points. 

Of course, many of our scan points have $m_h<2 m_S$, for which the mediator $h$ decay is off-shell. While the mono-jet cross section in this case is more involved (proportional to the square of the product of the $h$ couplings to DM and to SM fermions divided by the off-shell $h$ propagator), the current constraints in this regime from the LHC turn out to be extremely weak~\cite{Abdallah:2014hon}. 

%%%%%%%%%%%%%%%%%%%%%%%%%%%%%%%%%
\subsection {Dark matter direct detection}
%%%%%%%%%%%%%%%%%%%%%%%%%%%%%%%%%

\begin{figure}[t]
\begin{center}
\includegraphics[width=0.58\textwidth]{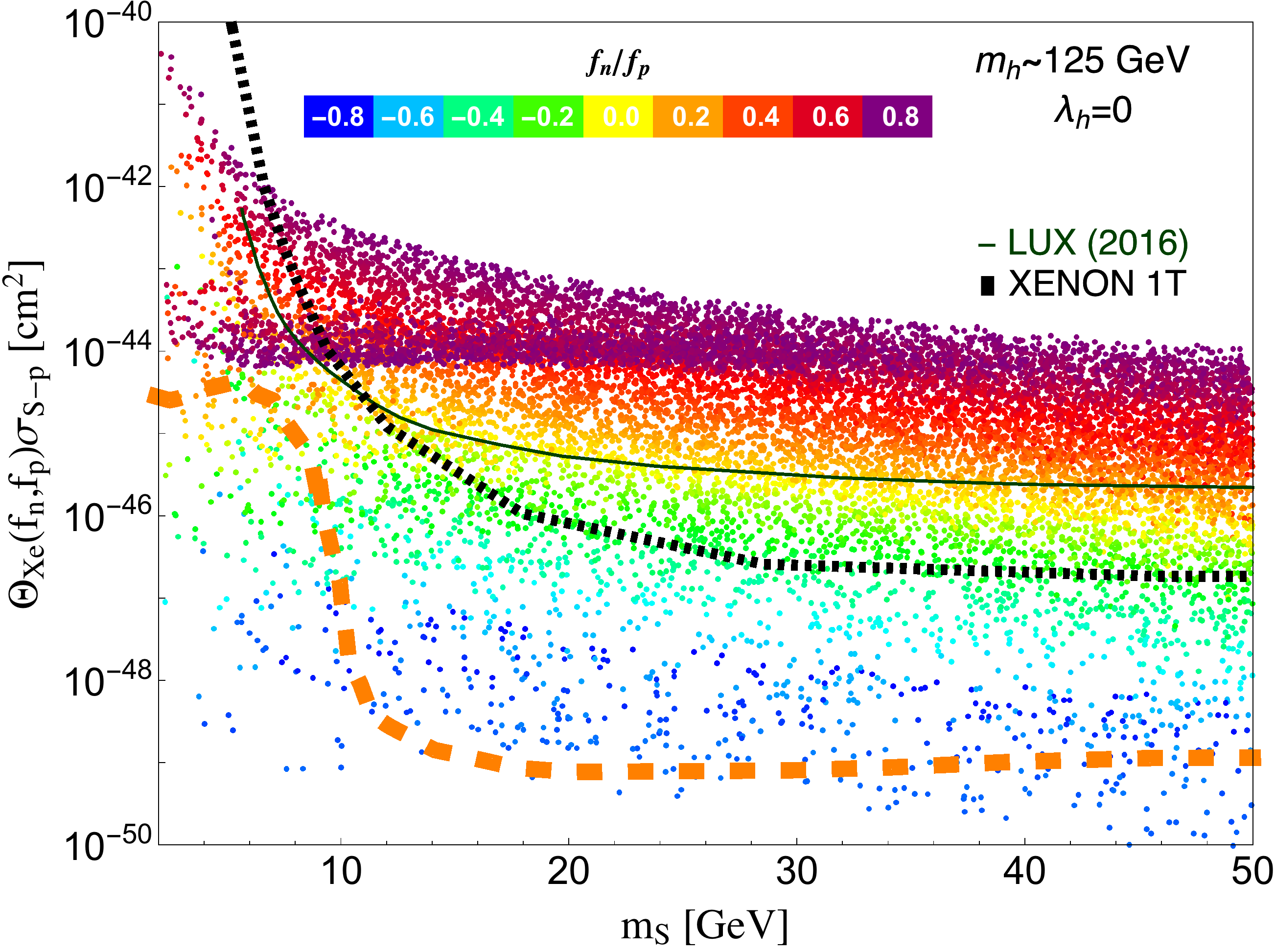}
\includegraphics[width=0.58\textwidth]{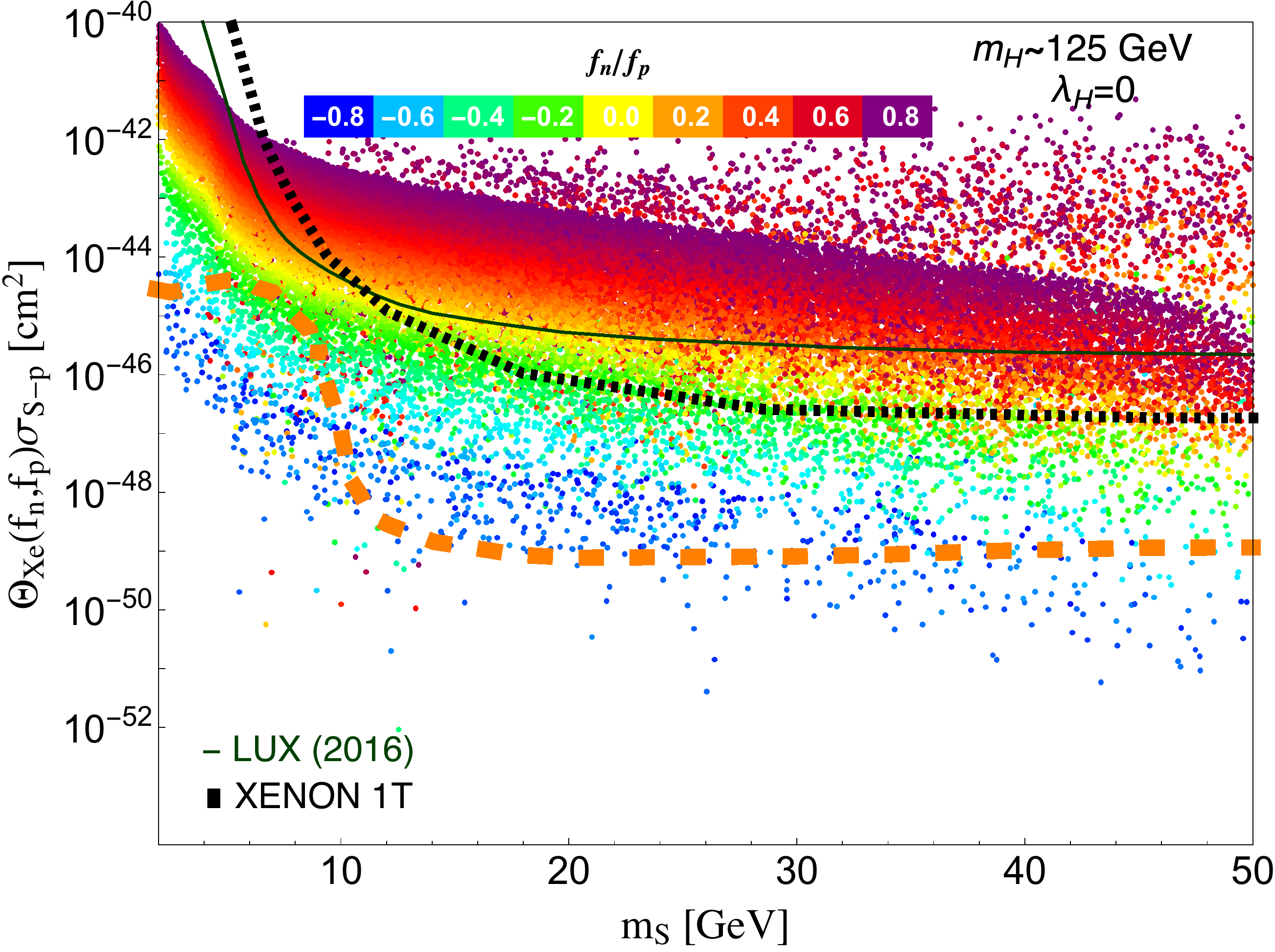}
\end{center}\vspace*{-5mm}
\caption{For points with correct $\omgs$, we show $\Theta_{\rm Xe} \sigma^{\rm SI}_{S-p}$ vs.  $m_S$ for the $h125$ (upper) and $H125$ (lower) cases compared to the LUX~(2016) bound~\cite{Akerib:2013tjd} (solid dark green) 
and the XENON1T (2017) projections (dark dashed green boxes)~\cite{Aprile:2012zx}.
All points shown satisfy the SuperCDMS limits.
The neutrino coherent scattering dominates the recoil spectrum below the thick dashed orange line.}
\label{mS-Xsec-fnp}
\vspace*{-.1in}
\end{figure}

In Fig.~\ref{mS-Xsec-fnp}, we show the expected cross sections for $S$ scattering off nuclei in Xenon-based detectors for both the $h125$ and $H125$ cases  
together with  LUX~(2016) results and the XENON1T future projection. The points are colored with respect to $\fnp$.
Note that, in accordance with expectations, points for which the cross section is suppressed correspond to $\fnp$ approaching $-0.7$.
The conclusion from the plots is that, after including isospin-violation, the 2HDMS  could easily be consistent with both the
LUX~(2016) limits and also the limits anticipated for XENON1T. 
Conversely, future improved exclusion limits or positive signals will either place an upper bound on $\fnp$ or favor a particular value of $\fnp$.

We have also examined the predicted cross sections for Si  and Ge detectors. For both the $h125$ and $H125$ scenarios, our points (which satisfy the SuperCDMS and LUX~(2016) limits) have cross sections at least two orders of magnitude below any of the tentative signals 
(CDMS-II~\cite{Agnese:2013cvt,Agnese:2013rvf}, DAMA~\cite{Bernabei:2013xsa}, CoGeNT~\cite{Aalseth:2010vx}, and CRESST-II~\cite{Angloher:2011uu}) found in the low mass region.

%%%%%%%%%%%%%%%%%%%%%%%%%%%%%%%%%%%%%%%%
\subsection {Dark matter indirect detection}
%%%%%%%%%%%%%%%%%%%%%%%%%%%%%%%%%%%%%%%%
\begin{figure}[t]
\begin{center}
\includegraphics[width=0.45\textwidth]{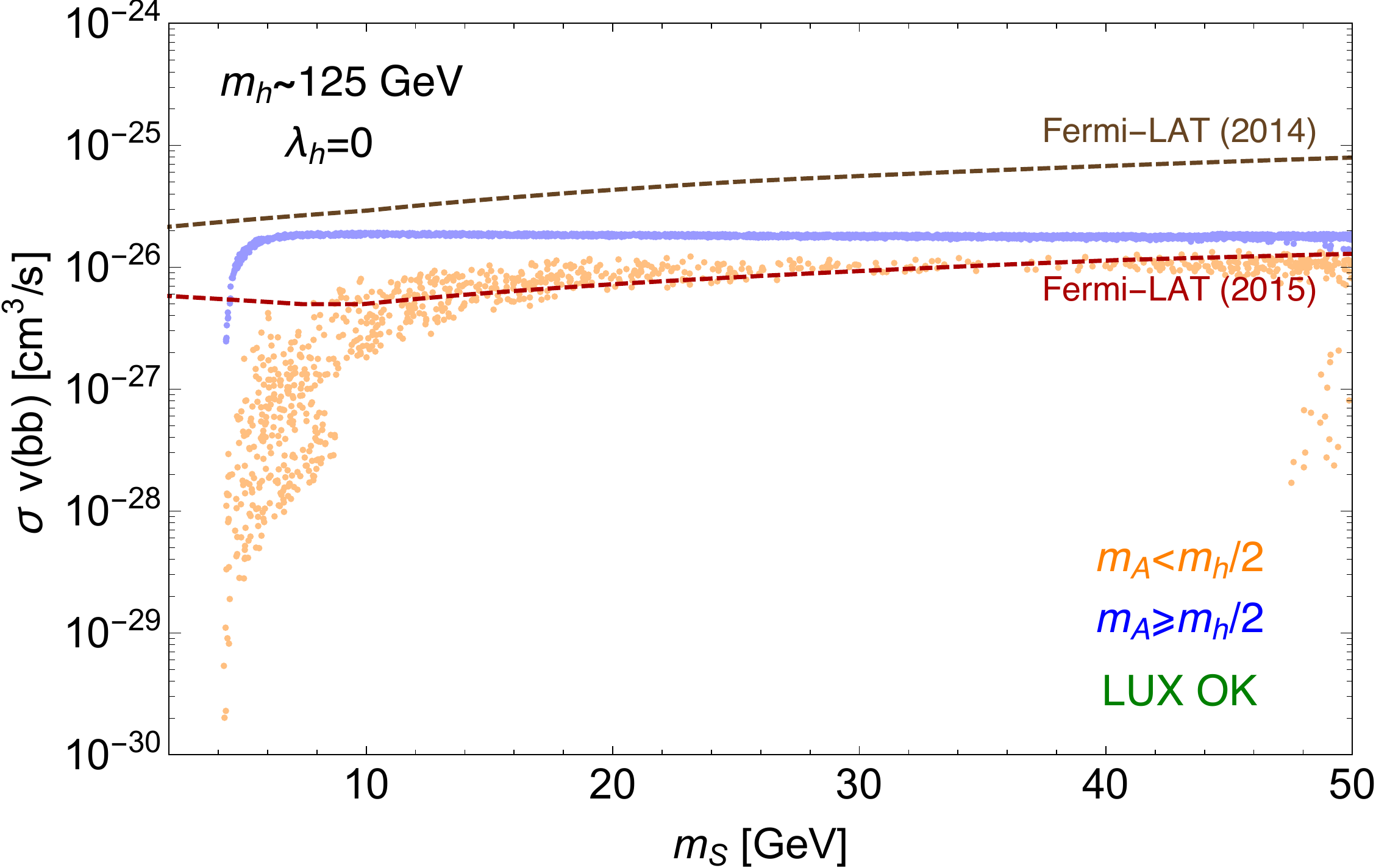}
\hspace*{-.06in}
\includegraphics[width=0.45\textwidth]{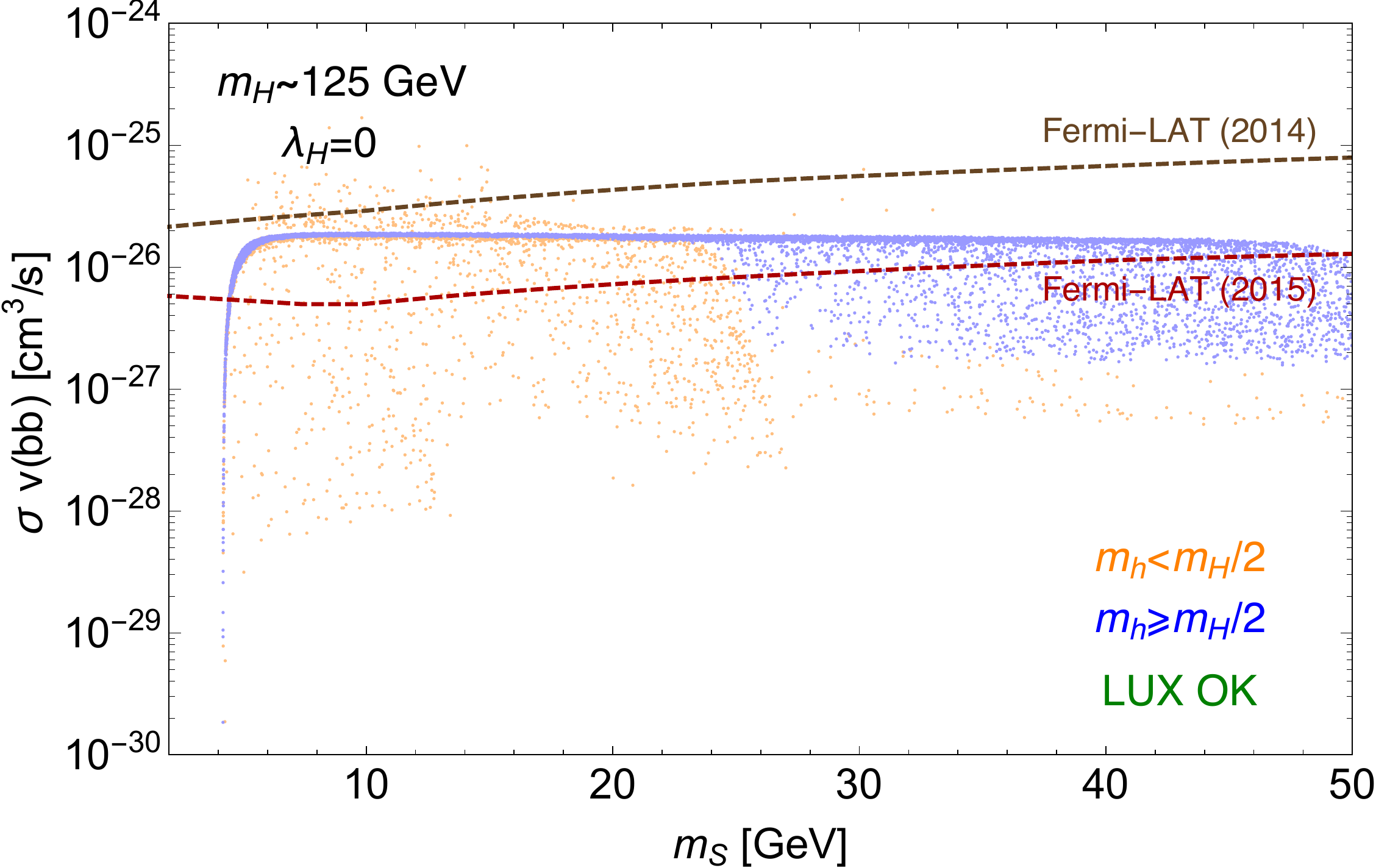}
\includegraphics[width=0.45\textwidth]{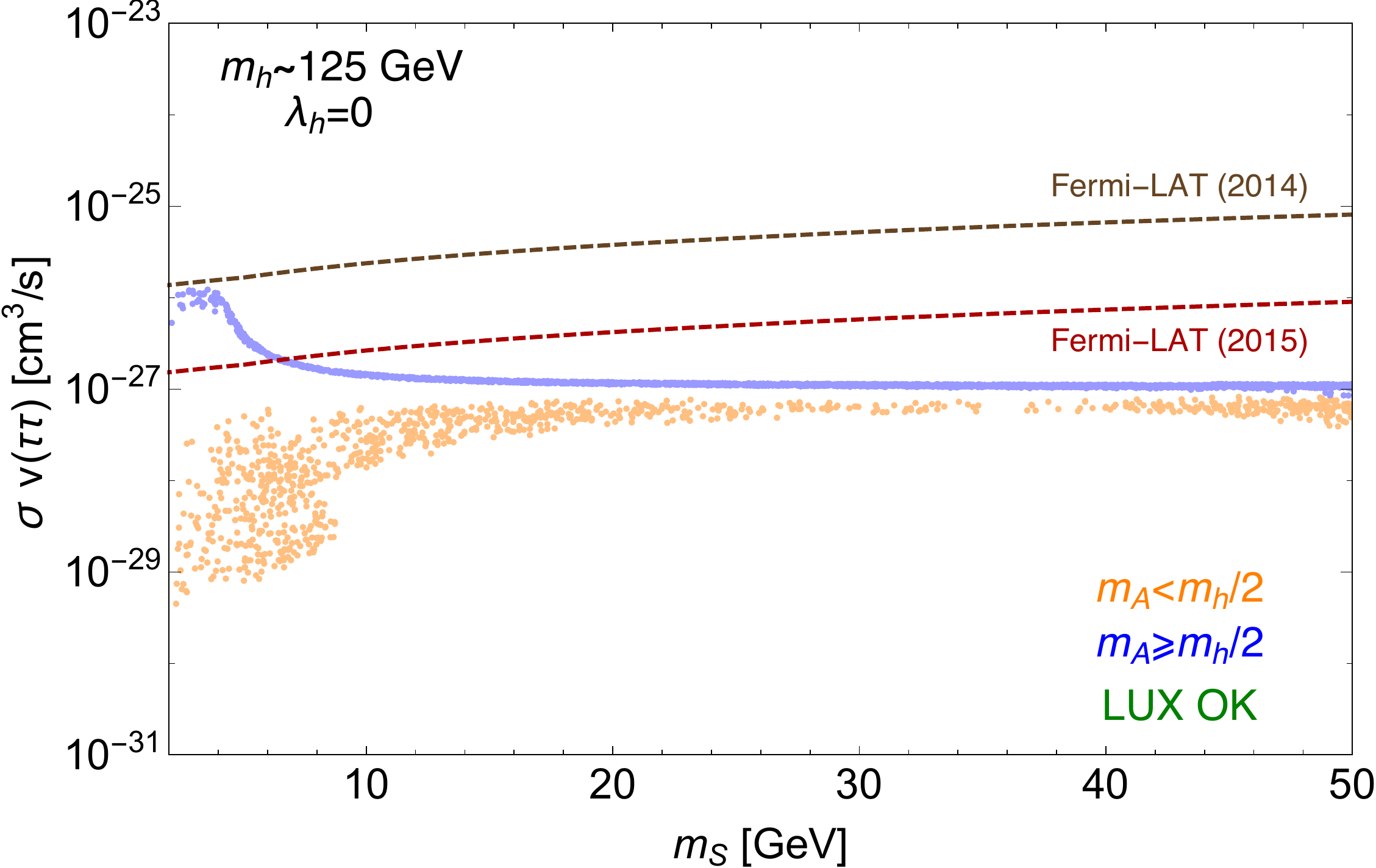}
\hspace*{-.06in}
\includegraphics[width=0.45\textwidth]{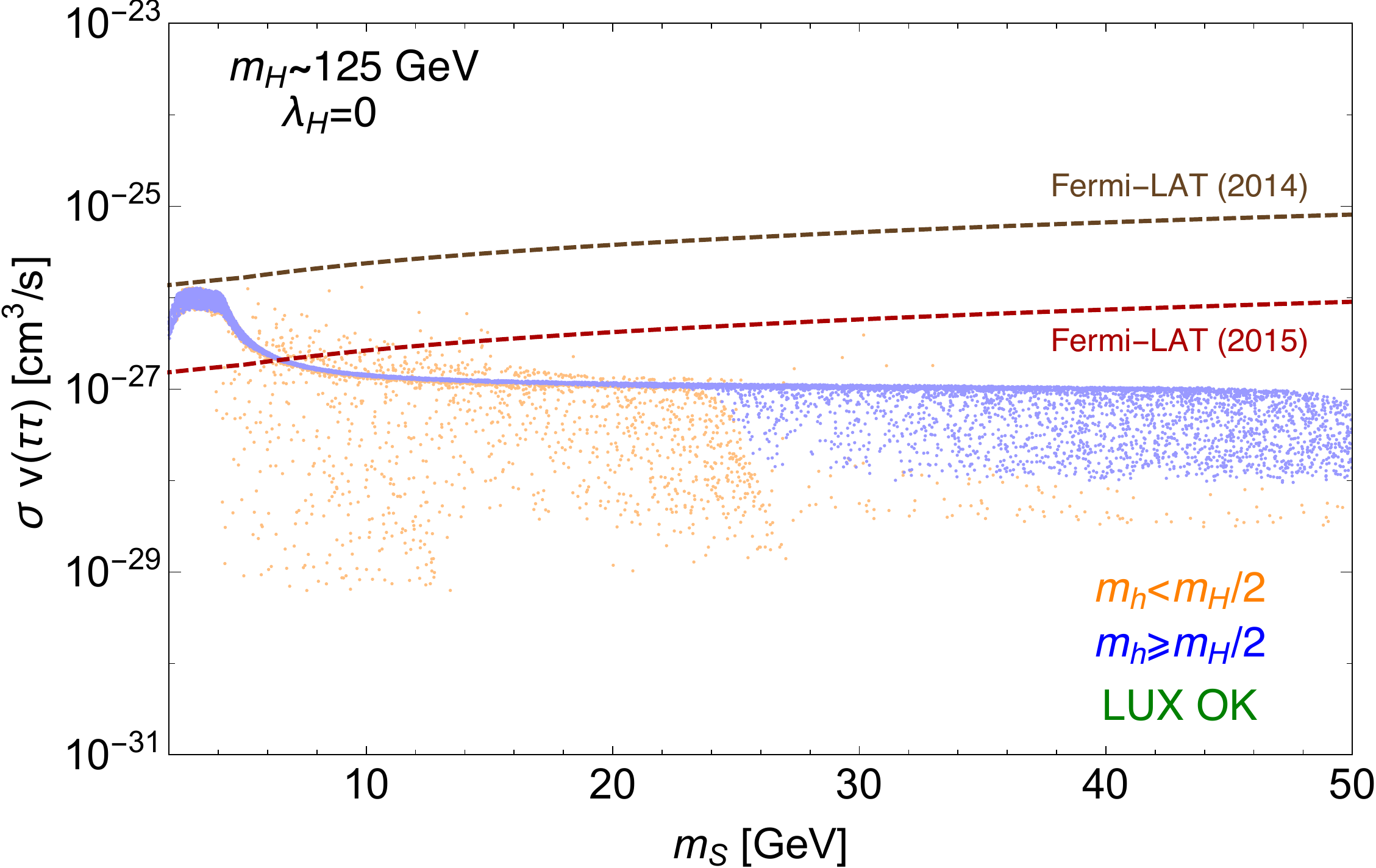}
\end{center}
\vspace*{-5mm}
\caption{Indirect detection cross sections for $m_h \sim 125 \gev$ (left) and $\mhh\sim 125\gev$ (right) compared to Fermi-LAT limits for $b\bar b$ and $\tau^+ \tau^-$ annihilations shown in the upper and lower panels, respectively. All points have correct $\omgs$ and obey the LUX~(2016) and SuperCDMS limits.}
\label{fig:ind}
\vspace*{-.1in}
\end{figure}

Finally, we consider the limits from 
indirect detection of $SS$ annihilation products. If DM annihilates it could produce pairs of SM particles, such as electron-positron pairs or photons. Currently, there are limits from the Fermi-LAT collaboration, see \cite{Ackermann:2013yva} and \cite{Ackermann:2015zua}, on this annihilation cross section coming from the observation of the dwarf spheroidal galaxies of the Milky Way, which are the most DM-dominated objects we know of. We do not consider limits related to the observation of the Galactic Center~\cite{Daylan:2014rsa} since they depend strongly on the choice of the DM profile. 

Our results for indirect detection related to the $b\bar b$ and  $\tau^+ \tau^-$ final states are shown in Fig.~\ref{fig:ind}. As described below, the $\tau^+ \tau^-$ final state must be considered for $m_S\leq m_b$.
In the $h125$ case, we observe that the points which survive the LUX~(2016) limits {\it and} obey the Fermi-LAT (2015) limits are those with $\mha < 62.5\gev$. 
Note that even a factor of 2 improvement in the Fermi-LAT limits would exclude all $h125$ points with $m_S\gsim 12\gev$. In the $H125$ case, we compare points with $\mhl<62.5\gev$ to points with $\mhl\geq 62.5\gev$.  Regardless of the $\mhl$ choice or the value of $m_S$, a large number of points survive the current Fermi-LAT limits and a significant fraction will also survive improved limits. 
In the $h125$ ($H125$) case, all the blue (all the) points shown below the $b\bar{b}$ threshold are  eliminated by the $\tau^+ \tau^-$ final state limits.

After all constraints, the LHC phenomenology of the non-SM-like 2HDM Higgs bosons  is easily summarized. First,  {\it all} must lie in definite mass ranges below $650\gev$. The allowed mass range for all scalar bosons in various scenarios is summarized in Table~\ref{tab:massummary}. 
Second, $\beta\sim \pi/4$ ($\tanb=1$) and $\alpha \sim -\pi/4~(+\pi/4)$ in the $h125$ ($H125$) \typeii\ scenarios  imply nearly unique Higgs-quark couplings. The resulting  direct production cross sections at 13 TeV for all these non-SM-like 2HDM Higgs bosons will be substantial. Further, their decays will be such that detection should be possible. In the $h125$ scenario, 
$H^\pm\to tb$ is always dominant ($\hpm\to H W^\pm$ is  kinematically forbidden) while $H \to SS, AZ, t\bar {t}$ constitute the main decays for the $H$. In the relevant range of $\mha\lsim 62.5\gev$, $A\to b\bar b~(\tau\tau)$ dominates for $\mha>2m_b$ ($\mha<2m_b$).
For the $H125$ scenario, 
the important modes are $\hpm\to h W^\pm,tb$ and $A\to Zh,t\bar t$. The $h$ will decay to a mixture of $b\bar b$ and $SS$ (invisible) final states. As an example, Ref.~\cite{Casolino:2015cza} claims that $t\bar t A$ production with $A\to b\anti b$ will be detectable at the LHC Run~2 for $\tanb=1$ if $\mha\in[20,100]\gev$.

\begin{table}[t]
\vspace{-.2in}
\caption{The allowed mass range for the scalars in various scenarios. The units are in GeV.}
\label{tab:massummary}
\begin{center}
\begin{tabular}{|c|c|c|c|c|c|}
\hline
Scenario & $m_S$ & $m_h$ & $m_H$ & $m_A$ &  $m_{H^\pm}$ \cr
 \hline
 $h125$ & $\lesssim 12$ & 125   &  $ 440-650$  & $ \lesssim 62.5 $ & $ 485-630 $   \cr
\hline
 $H125$ &  $\gsim 4$ & $ 10-62.5 $  &  125 & $ 420-650$  & $ 485-630$  \cr
 \hline
 $H125$ & $\gsim 25$ & $ 62.5-125 $  &  125 & $ 420-650 $ & $ 485-630$  \cr
\hline 
\end{tabular}
\end{center}
\vspace{-.15in}
\end{table}

%%%%%%%%%%%%%%%%%%%%%%%%%%%%%%%%%%%%%%%%
\vspace*{-.1in}
\section{Conclusions}
\vspace*{-.1in}

In a multi-Higgs model in which one Higgs fits the LHC $125\gev$ state, one or more of the other Higgs bosons can mediate DM-nucleon interactions.  We have shown that for appropriate Higgs-quark couplings maximal DM isospin violation is possible  independent of the nature of DM.  We then considered the explicit example of a \typeii\ 2HDM where the $h$ ($H$) is identified with the LHC $125\gev$ state while the $H$ ($h$)  mediates the coupling between quarks and DM. This allows us to have DM of correct relic density that can even be maximally isospin violating (for 2HDM parameters $\tanb\sim 1$ and $\alpha \sim \pm \pi/4$), thereby evading LUX~(2016) and future XENON1T limits even at low DM mass. If DM is discovered in the future, then the level of the observed direct detection cross section will determine the $\fnp$ value and the relevant $\tanb$ and $\alpha$ which can, hopefully, be checked against direct Higgs sector observations. 

We next considered the 2HDMS model in which a scalar singlet, $S$, is added to the 2HDM, showing that it can be a viable DM particle in both the $h125$ and $H125$ scenarios.  In the former (latter), the $hSS$ ($HSS$) coupling can be sufficiently suppressed that the $S$ does not affect the purely 2HDM fits of the $h$ ($H$) to the $125\gev$ signal, while the $HSS$ ($hSS$) coupling can be chosen to give correct $\omgs$.  By employing appropriate isospin-violating 2HDM parameters, one can avoid direct and indirect detection limits even at low $m_S$. In this model, the non-SM-like Higgs bosons will be discovered during LHC Run~2 due to the fact that their masses and couplings are strongly restricted.

It is also worth mentioning that the single DM scalar scenario of the 2HDMS considered here 
can be easily extended to a multi-component DM sector with $N$ real $O(N)$-symmetric scalars 
in the spirit of \cite{Drozd:2011aa}.

%%%%%%%%%%%%%%%%%%%%%%%%%%%%%%%%%%%%%%%%
\vskip -.25in
\section*{Acknowledgments} 
\vskip -.1in
JFG and YJ acknowledge partial support by US DOE grant DE-SC-000999. 
YJ also received generous support from LHC-TI fellowship US NSF grant PHY-0969510 and the Villum Foundation.
He also acknowledges the LATPh for hospitality and particularly thanks Genevi\`eve B\'elanger for useful discussion and technical assistance regarding micrOMEGAs. The authors are also grateful to Yushin Tsai and Jian Wang for their helpful discussing the mono-jet search for dark matter at the LHC. The work of AD was supported by the STFC Grant ST/J002798/1. BG acknowledges partial support by the National Science Centre, Poland decision no DEC-2014/13/B/ST2/03969 and DEC-2014/15/B/ST2/00108. JFG acknowledges support during revision while at the Aspen Center for Physics, which is supported by National Science Foundation grant PHY-1066293.

%%%%%%%%%%%%%%%%%%%%%%%%%%%%%%%%%%%%%%%%
\bibliographystyle{JHEP}
\vskip -.3in
\bibliography{IVDMrev}
%%%%%%%%%%%%%%%%%%%%%%%%%%%%%%%%%%%%%%%%

\end{document}